\documentclass[a4paper,11pt]{article}
\usepackage{amsmath, amssymb}
\usepackage{aaskaiid}
\usepackage{aas_macros}
\usepackage{threeparttable}
\usepackage{orcidlink}
\usepackage{rotating}

\title{Radio emission from ultra-diffuse galaxies residing in galaxy clusters}
\ShortTitle{Ultra-diffuse galaxies}

\author[1]{Dharam V. Lal\orcidlink{0000-0001-5470-305X}}
\ShortName{Lal \& Jeyasiona} 
\author[2]{Jeyasiona Murugesh}

\affiliation[1]{National Centre for Radio Astrophysics - Tata Institute of Fundamental Research, Post Box 3, Ganeshkhind P.O., Pune 411007, India}
\emailAdd{dharam@ncra.tifr.res.in}
\affiliation[2]{Department of Agricultural and Food Engineering, Indian Institute of Technology 
Kharagpur, Kharagpur 721302, India}

\abstract{
Ultra-diffuse galaxies (UDGs), defined by their extremely low surface brightness ($g$-band $\mu \gtrsim 24$ mag arcsec$^{-2}$) and large effective radii (3--10 arcsec), remain one of the most puzzling galaxy populations in the nearby Universe \citep{vanDokkum2015}. Predominantly found in dense environments, UDGs in the Coma cluster show a preferential alignment of their major axes toward the cluster centre, suggesting strong environmental influence on their formation and evolution. Using high-sensitivity, low-frequency radio data from the upgraded Giant Metrewave Radio Telescope (GMRT), a pathfinder instrument we examined all 854 UDGs cataloged in Coma cluster \citep{Yagi2016}. Despite the unprecedented depth of these observations, no individual detections were made. A median stacking analysis in the upgraded GMRT band-3 achieved a 5 $\times$ \textsc{rms} upper limit $\simeq$1.5~$\mu$Jy, providing the most stringent constraint yet on the average (mean) radio emission from UDGs, corresponding to star-formation rates $\lesssim$10$^{-3}$~M$_\odot$~yr$^{-1}$ for Coma-cluster-like systems and $\lesssim$10$^{-1}$~M$_\odot$~yr$^{-1}$ at $z \sim 0.05$.

Looking ahead, the Square Kilometre Array (SKA) will transform the study of such faint galaxies. While the early AA$^\star$ configuration will deliver sensitivities comparable to the upgraded GMRT, the AA4 design baseline will achieve sub-$\mu$Jy \textsc{rms} levels at matched frequencies ($\nu \sim 200$~MHz--1.4~GHz), enabling detections of UDGs with star formation rates as low as 10$^{-4}$--10$^{-3}$~M$_\odot$~yr$^{-1}$ within Virgo and Coma distances. Such capabilities will allow robust discrimination between quenched, dark-matter-dominated systems and those sustaining weak residual star formation or low-luminosity nuclear activity.

Clearly, the forthcoming sensitivity of SKA-Mid and SKA-Low will be decisive in determining whether UDGs extend the faint tail of the active galaxy or star-formation luminosity function, or instead constitute a distinct radio-silent class. Even if the future looks dark for UDGs, in the sense that their radio emission may remain extremely challenging to detect, there is still good hope that upcoming facilities will possibly reveal their true nature.}

\begin{document}
\maketitle

\section{Introduction}

Ultra-diffuse galaxies (UDGs) are an extreme population of low-surface-brightness systems, especially abundant in rich galaxy clusters. \citet{vanDokkum2015} defined them as galaxies with central $g$-band surface brightness $\mu_g \gtrsim 24$~mag~arcsec$^{-2}$ and large effective (half-light) radii $R_e \gtrsim 1.5$~kpc \cite[see][for an in-depth discussion on UDG definition]{VanNest2022}.
Thousands of UDGs have now been identified across a wide range of environments, e.g.,
Abell\,2744: \cite{Janssens2019}, Coma cluster: \cite{Yagi2016}, \cite{Zaritsky2019,Zaritsky2021}, Globular Cluster system: \cite{Gannon2024}, Pisces-Perseus super-cluster: \cite{Martinez2016}, Perseus cluster: \cite{Gannon2022}, Stripe~82: \cite{Barbosa2020}, UGC 6594 group: \cite{Gannon2021}, Virgo cluster: \cite{Mihos2015}, etc.
Observed cluster UDGs span a substantial luminosity range, an absolute magnitude of about $-12$ to $-16$ \citep{Koda2015}, although the completeness at the faint end of their luminosity function remains uncertain. Recent studies suggest that more than 7\% of all galaxies may be ultra-diffuse \citep{Li2023}.
Understanding how these galaxies form and evolve has therefore become a challenge in galaxy formation research, with UDGs emerging as valuable laboratories for studying both galaxy formation and cluster evolution.

A wide variety of theories have been proposed to explain the formation of UDGs, generally invoking either (i) external processes such as tidal heating, ram-pressure stripping, environmental quenching, or galaxy mergers \citep[][etc.]{Carleton2019,Sales2020,Jones2021,vanDokkum2022}, or (ii) internal mechanisms such as high dark-matter halo spin, stellar feedback, or passive stellar evolution \citep[][etc.]{DiCintio2017,Benavides2023,Fielder2024}. In many cases, a combination of both internal and external factors possibly contributes to their formation \citep[see also][and references therein]{Martin2019,Gannon2024}, and different formation mechanisms are expected to leave distinct signatures in the stellar populations and dark matter halos of UDGs. For instance, UDGs formed through episodic stellar feedback should exhibit extended star formation histories, low metallicities, and dwarf-galaxy-like halos with correspondingly low velocity dispersions and few globular clusters \citep{Zheng2024}. In contrast, UDGs that formed early and quenched rapidly are expected to host old, metal-poor stellar populations and reside in more massive halos, with higher velocity dispersions and richer globular-cluster-like systems \citep{Gannon2024}.

Observational studies have shown a transition in the overall properties of field and cluster UDGs, with UDGs in the field tending to be bluer, more irregular, and gas-rich, with some ongoing star formation, while those in clusters are mostly red, spheroidal, and quenched \citep[e.g.][]{vanderBurg2016,Roman2017,Roman2017a}.
Although most field UDGs are gas-rich and star-forming, a population of isolated quenched systems has also been reported, which may originate as backsplash galaxies that were previously satellites of groups or clusters, as suggested by simulations \citep{Benavides2021}.
Large optical surveys and dedicated catalogs \cite[e.g.,][]{Yagi2016} have revealed that clusters such as Coma host hundreds of UDGs, far exceeding field expectations, and that their structural properties (e.g. axis ratios, sizes) are remarkably coherent, suggesting that environmental effects play a significant role in shaping them \citep{Motiwala2025}. Notably, (i) many UDGs appear as barely resolved diffuse systems along the line of sight to the Coma cluster \citep[$R_e \sim$ 800 pc–-5~kpc, effective surface brightnesses = 25--28 mag~arcsec$^{-2}$ and stellar masses $\sim$ 1 $\times$ 10$^{7}$ M$_\odot$ -- 5 $\times$ 10$^{8}$ M$_\odot$,][]{Koda2015}, and (ii) deep stacked Subaru images show little evidence of strong tidal disturbances in their outskirts, implying either substantial dark-matter dominance or survival mechanisms that shield them from rapid disruption \citep{Yagi2016}.

Results from cosmological simulations, e.g., Millennium-II and Phoenix: \cite{Rong2017}, RomulusC: \cite{Tremmel2020}, TNG50: \cite{Benavides2023}, EAGLE: \cite{Zheng2025}, NIHAO: \cite{Motiwala2025}, etc., present a picture in which UDGs are not a wholly separate galaxy class, but rather the extreme tail of the dwarf–galaxy population emerging under particular conditions. Many simulated UDGs occupy dwarf-galaxy-like mass halos (M$_\star \sim 10^7$--$10^9$ M$_\odot$), and yet are unusually extended (effective radii $\gtrsim$ 2--3 kpc) because of high halo spin, early quenching, or tidal/heating processes. In simulations focusing on clusters of galaxies \citep{vanDokkum2022}, UDG analogues often fall into massive halos early and suffer strong environmental processing, e.g., ram-pressure stripping, tidal mass loss and adiabatic expansion, leading to low surface brightness and gas-poor states \citep{Tremmel2020}.  Alternatively, in field environments, simulations show a subset of ``born UDGs'' whose gas spins and feedback‐driven expansions yield large radii without heavy stripping. Across environments the result is that UDGs tend to have large sizes for their stellar mass, older stellar populations, and a continuous rather than discrete distribution with respect to normal dwarf galaxies \citep{Benavides2023}.
Thus, understanding UDGs in clusters is important because their origin, whether they are ``puffed-up'' dwarfs, failed milky-way-like galaxies, or a mix of both, which contains key information about galaxy formation at low surface brightness and the environmental processes that quench star formation. Their abundance scales with the mass of their host halo, makes them valuable tracers of dark-matter-dominated regions and environmental processes during the formation and growth of galaxy clusters that affect their member galaxies.

Although optical surveys have established large samples of UDGs across a range of environments, only a few attempts have been made to detect them at radio wavelengths. Early efforts targeted their H\,I and continuum emission to probe gas content and star formation activity. \cite{Papastergis2017a} used the Effelsberg 100\,m telescope to search for H\,I in four isolated UDGs, detecting one (SdI--2) with M$_{\mathrm{H\,I}} \approx 2.4 \times 10^8$\,M$_\odot$, while setting upper limits of $(1.3$--$2.4)\times10^8$\,M$_\odot$ for the others. \cite{Scott2021} presented resolved H\,I observations of two blue, gas-rich (isolated) UDGs using the GMRT, confirming that such systems exist outside clusters but differ from the quiescent, red UDGs found in dense environments. In contrast, \cite{Struble2018} searched for radio counterparts to UDGs in the Coma cluster using VLA data and found no significant matches above background levels, indicating that cluster UDGs are largely radio quiet. More recently, the SMUDGes H\,I survey \citep{Karunakaran2024} reported detections for 110 out of 378 UDG candidates, predominantly in isolated environments, again confirming that gas-rich UDGs are primarily field systems while cluster UDGs are strongly quenched.
This motivates deep, targeted radio continuum and H\,I observations of UDGs using current and upcoming facilities.
Pathfinder and precursor arrays (upgraded GMRT, MeerKAT, LOFAR), with their exceptional sensitivity and wide-field imaging capabilities, are ideally suited to bridge the gap between deep optical surveys and the unprecedented resolution and sensitivity that the SKA will deliver. Such observations will enable the first statistically significant constraints on the non-thermal emission, magnetic fields, and cold-gas reservoirs of UDGs across a range of environments.

\section{First results from upgraded GMRT}

\begin{figure}
\begin{center}
\begin{tabular}{c}
   \includegraphics[width=14.5cm]{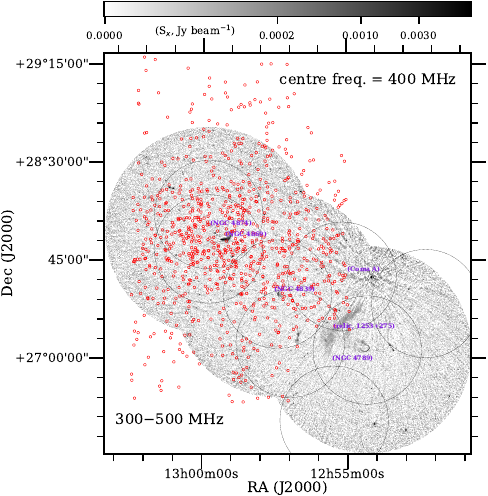}
\end{tabular}
\end{center}
\caption{Mosaic image of three pointings in the upgraded GMRT band-3 (250-–500 MHz) at an angular resolution of $\sim$6.5$^{\prime\prime}$.
UDG positions from from \citet{Yagi2016} are marked by the ``{\color{red}{$\circ$}}'' sign.
The gray-scale image is displayed in logarithmic scales to emphasize the extended, low-surface brightness diffuse radio emission, and the names of key sources are labeled.  The \textsc{rms} noise is relatively uniform between 21 $\mu$Jy~beam$^{-1}$ and 31 $\mu$Jy~beam$^{-1}$, apart from small regions near bright sources, e.g., Coma A where the \textsc{rms} noise increases. Eight circles correspond to the eight pointings in the 550--850 MHz band of upgraded GMRT.
Note the starry-pattern artifact centered on Coma A, limiting the dynamic range.}
\label{fig:fig-1}
\end{figure}

Although pathfinder and precursor facilities have achieved remarkable sensitivity to diffuse cluster emission, direct radio detections of individual UDGs remain rare.
The UDG sample is drawn from the catalog of \citet{Yagi2016}, who identified 854 such systems in the Coma cluster (z = 0.02316) through a systematic search of deep $R$-band Suprime-Cam/Subaru imaging. Their study provides a comprehensive and homogeneous compilation of UDGs in a dense cluster environment, forming a well-defined sample for statistical investigations.
We use this optically selected catalog as the basis for a radio study, aiming to examine the presence and properties of faint radio emission from these galaxies and to explore their contribution to the low-luminosity end of the radio population.

We therefore analyzed the Coma cluster UDGs using archival low-frequency radio continuum observations obtained with the upgraded Giant Metrewave Radio Telescope (GMRT), a pathfinder instrument. The UDGs lie within the area covered by the deep upgraded GMRT survey of the Coma cluster \citep{Lal2022}. We use observations in both band~3 (250–500~MHz) and band~4 (550–850~MHz), which together provide high-sensitivity, arcsecond-scale resolution across the cluster environment.
These final images have a \textsc{rms} noise level of 21.1$-$36.6~$\mu$Jy~beam$^{-1}$ and 12.8$-$42.4~$\mu$Jy~beam$^{-1}$ at the angular resolutions of $\sim$6.1$^{\prime\prime}$ and $\sim$3.7$^{\prime\prime}$ in the 250--500 MHz and 550--850 MHz bands, respectively.
After matching the two images at an angular resolution of $\sim$6.5$^{\prime\prime}$, we extracted a region of 50$^{\prime\prime}$ $\times$ 50$^{\prime\prime}$ from these band-3 and band-4 images centered on the UDG position as given by the Subaru survey for each UDG in our sample.
We retained 736 and 582 UDGs suitable for stacking in band~3 and band~4, respectively.
Figure~\ref{fig:fig-1} shows the mosaic image of three pointings in the upgraded GMRT band-3 (250–500 MHz) at an angular resolution of $\sim$6.5$^{\prime\prime}$. Eight circles correspond to the eight pointings in the 550--850 MHz band of upgraded GMRT. UDG positions are marked by the ``{\color{red}{$\circ$}}'' sign.

Surprisingly, no individual UDG was detected in either band above the $3\sigma$ (= 3 $\times$ \textsc{rms}) threshold within the sensitivity limits of the images. Thus, we explored two complementary methods for stacking radio images, mean and median stacking, each with distinct advantages and limitations \citep{White2007}. Mean stacking provides an easily interpretable measure of average flux density but is highly sensitive to outliers, such as bright sources or noisy images within the stack, which can bias the result. Although such effects can be mitigated by excluding sources above certain flux density or \textsc{rms} thresholds within the extracted region, the result remains sensitive to the specific cutoff criteria. Median stacking, on the other hand, is considerably more resilient to outliers and non-Gaussian noise distributions (e.g., \citealt{Gott2001}), allowing all images to be included without arbitrary truncation. However, at low signal-to-noise ratios, interpreting the median becomes less straightforward because the recovered value can be shifted toward the local mean, with the magnitude of the shift depending on the noise level. Thus, while the median provides stability against contamination, it can underestimate the true flux density for faint noise-dominated samples.

\begin{figure}
\begin{center}
    \begin{tabular}{cc}
   \includegraphics[height=6.2cm]{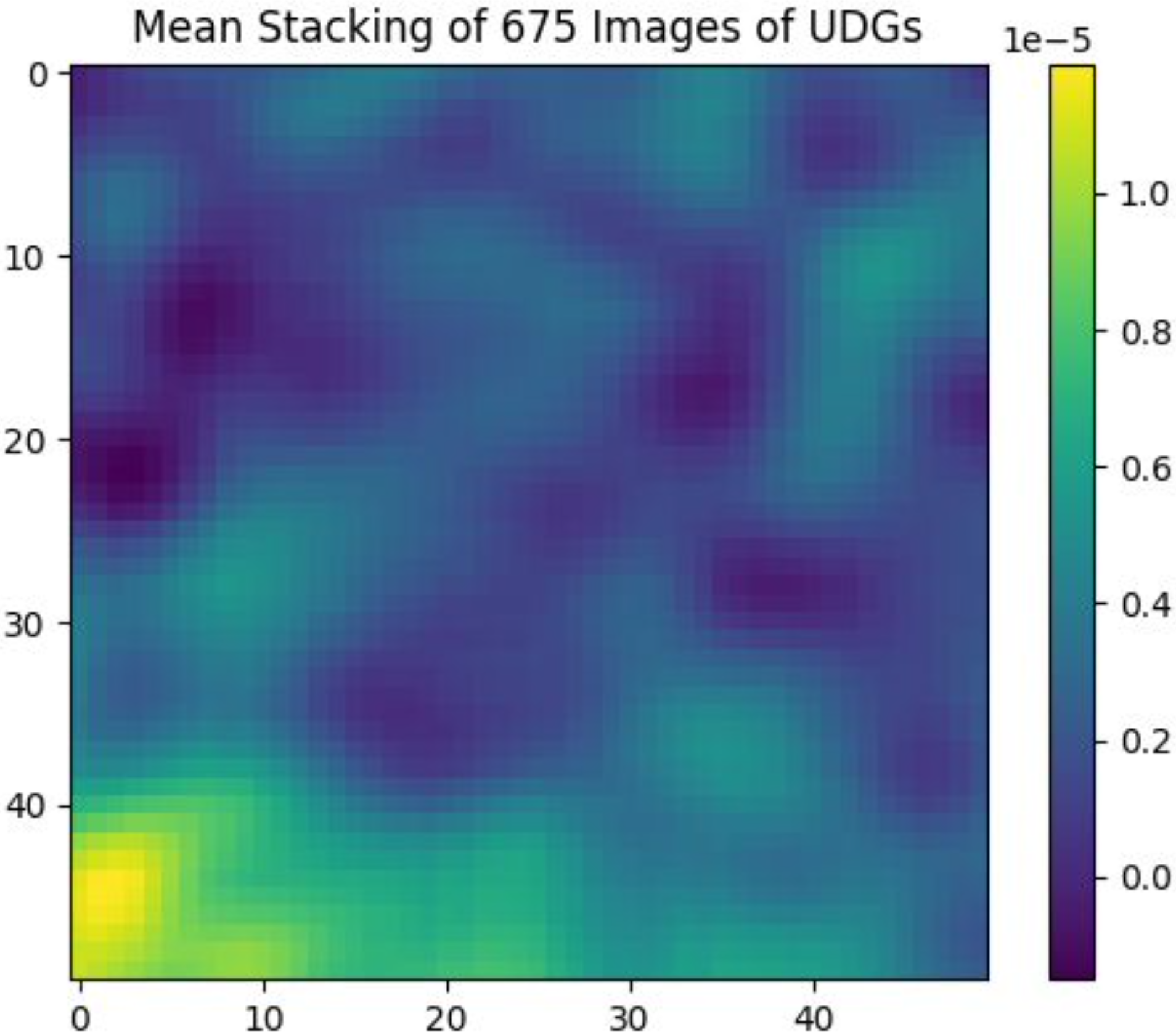} &
   \includegraphics[height=6.2cm]{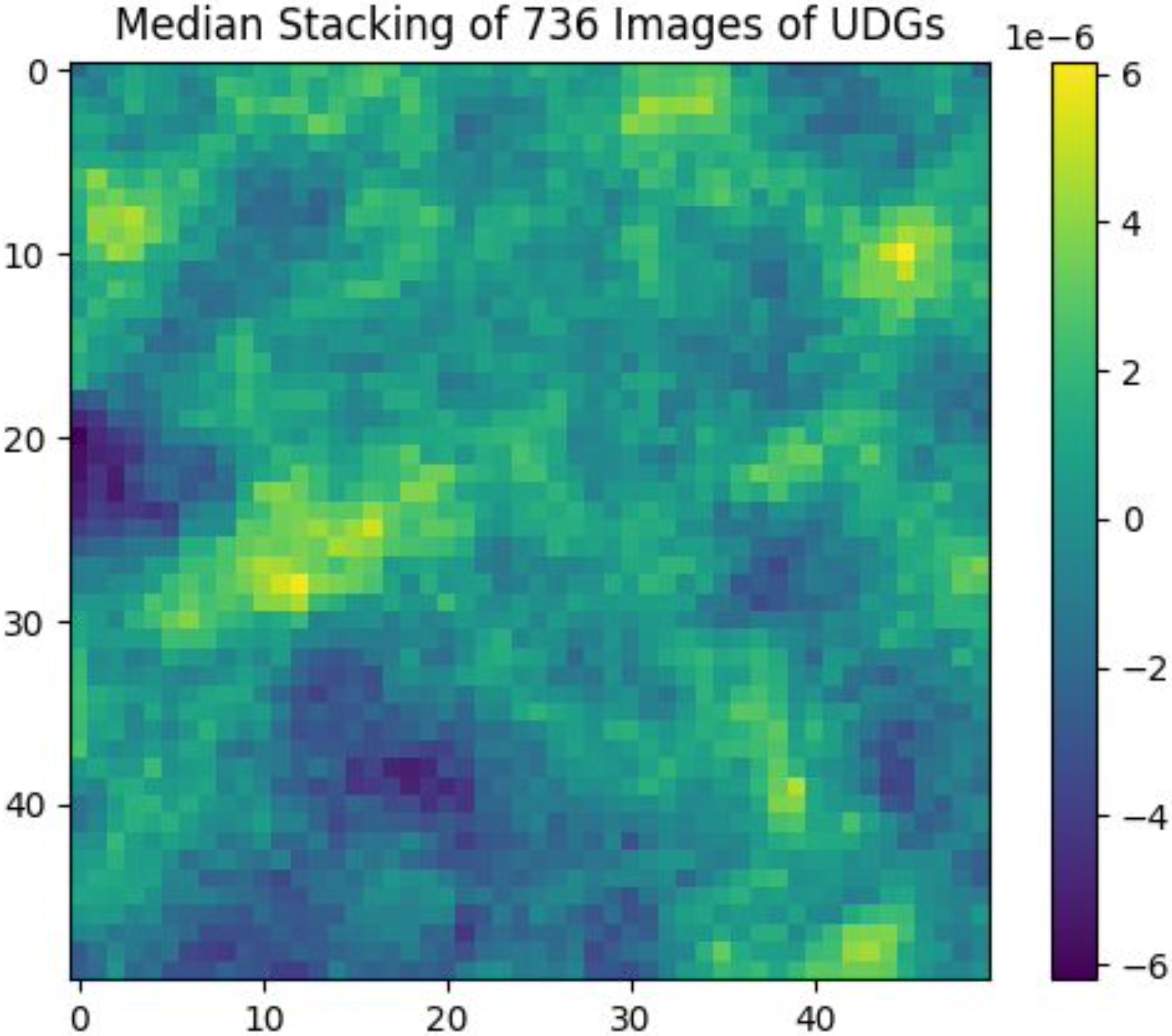}
    \end{tabular}
\end{center}
    \caption{Result of constructing a mean (left panel) and a median (right panel) stack of the 675 and 736, respectively source positions in the UDG catalog. The image displays a 50$^{\prime\prime}$ $\times$ 50$^{\prime\prime}$ color-scale image (pixel size = 0.8$^{\prime\prime}$).  The flux densities from the stacked images are consistent with the expected noise reduction ($\approx N^{-1/2}$), where $N$ is the number of images in the stack, confirming that the results are dominated by Gaussian noise rather than faint unresolved sources.}
    \label{fig:stack-images}
\end{figure}

To investigate the ensemble radio properties of these ultra diffuse galaxies, we performed mean and median stacking analyses in both bands, which offers a valuable cross-check, ensuring a more reliable characterization of the statistical radio properties. In band~3, we stacked 675 regions with no outliers for mean-stacking and all 736 images for median-stacking, whereas in band-4, we stacked 533 regions with no outliers for mean-stacking and all 582 images for median-stacking.
We did not detect any signal at the locations of the UDGs in any of the stacked images.  In band-3, the stacked image yielded an average flux density upper limit of $\sim$2.3~$\mu$Jy for mean-stacking and $\sim$1.7~$\mu$Jy median-stacking. Whereas in band~4, the stacked image yielded an average flux density upper limit of $\sim$1.6~$\mu$Jy for mean-stacking and $\sim$1.5~$\mu$Jy median-stacking.
Figure~\ref{fig:stack-images} shows resulting image at band-3 (= 50$^{\prime\prime}$ $\times$ 50$^{\prime\prime}$) of constructing a mean (left panel) and a median (right panel) stack of the 675 and 736, respectively source positions in the UDG catalog.  The resulting band-4 stacked images are (again non-detection and are therefore) not shown.
The flux densities measured from the stacked images are consistent with the expected noise reduction ($\approx {N}^{-1/2}$, where $N$ is the number of images in the stack), confirming that the results are dominated by Gaussian noise rather than faint unresolved sources.
We have also analyzed LOFAR LoTSS DR2 data \citep{LoTSS-DR2-2022} of the Coma cluster, but the prominent diffuse emission from the Coma radio halo limits our ability to reach deeper sensitivities through stacking.

We note that scaling the band-3 (central frequency = 400 MHz) limits assuming a spectral index of $\alpha = -0.7$ yields a limit of $\sim$1.1 $\mu$Jy at band-4 (central frequency = 700 MHz), which is about a factor of 1.4 higher than the current best estimates ($\simeq$1.5~$\mu$Jy) derived from the stacked images.
Thus, our median-stacked image at band-3 is the deepest among all data-sets, providing the most stringent observational limits on the average radio emission from UDGs.  Consequently, we adopt this stack for all subsequent analyses. The combination of broader bandwidth, better sensitivity, and a larger number of contributing sources makes the band-3 result particularly robust for constraining faint, diffuse synchrotron emission in these ultra-diffuse systems.

\paragraph{Upper limit on the star-formation rate}
Most of the non-thermal radio emission from galaxies arises from electrons that have diffused away from their acceleration sites in supernova remnants into quiescent regions void of current star formation, and using \citet[Eq.~17,][]{Murphy2011}, we determine star formation rate (SFR),
\[
\begin{aligned}
\mathrm{SFR}\ (\mathrm{M_\odot\,yr^{-1}}) &= \frac{L_{400}\left(\frac{1.4\ \mathrm{GHz}}{0.4\ \mathrm{GHz}}\right)^{\alpha}}{1.57\times10^{21}\ \mathrm{W\ Hz^{-1}}}, ~{\rm and} \\[3pt]
L_{\nu} \;(\mathrm{W\,Hz^{-1}}) &= 4\pi D_{L~(\rm m)}^2 \,(S_{\nu~(\mu\mathrm{Jy})}\times 10^{-32}); 
\end{aligned}
\]
\noindent where $D_L$ is the luminosity distance (= 3.086 $\times$ 10$^{24}$ m), and we assume
$\alpha=-0.7$.
These imply that the mean SFR of UDGs in Coma cluster is $\approx$ 4.8 $\times$ 10$^{-4}$~M$_\odot$~yr$^{-1}$, well below 10$^{-3}$~M$_\odot$~yr$^{-1}$, consistent with their quiescent optical colors and lack of individual radio counterparts.
Together, these findings support that UDGs in rich clusters are strongly quenched systems, possibly shaped by the dense intra-cluster medium and environmental stripping processes.
At present, these constitute the best observational limits on our understanding of UDGs.
This motivates deep, targeted radio continuum and H\,I observations, especially with the SKA, to constrain their cold-gas reservoirs, star-formation and active galactic nuclei activity,
and to probe how cluster environments shape and regulate the formation and survival mechanisms of these extremely diffuse galaxies.

\section{Predictions for SKA AA$^\star$ and AA4}

We note that the AA4 (design baseline) configuration forms the reference \citep{bourke2015advancing,Braun2015aska.confE.174B}. Specifically, the SKA-Mid array comprises 197 dishes, 133 15-m SKA dishes and 64 13.5-m MeerKAT dishes (with baselines out to 150 km), while the SKA-Low array consists of 512 low-frequency aperture array stations of 256 antennas each (with a maximum baseline of around 75 km).
On the other hand, AA$^\star$ is a reduced-scale, interim deployment by the SKA Observatory \citep[][and talk by Shari Breen, Head of Science Operations, 2025 October 20, ``Meeting with the SKAO and Book Chapters'']{Braun2024_SKAO_TEL_0000818}.
Despite its smaller collecting area and correspondingly lower sensitivity compared to AA4, AA$^\star$ will support the full range of observing modes, e.g., up to 16 subarrays, commensal operations, and a wide frequency coverage from 50 MHz to 15.4 GHz. In this configuration, SKA-Mid includes 144 dishes (with a maximum baseline of 36 km), and SKA-Low consists of 307 stations (with a maximum baseline of 74 km).

\paragraph{Expected \textsc{rms} sensitivities} Thus, the interim AA$^\star$ configuration and later the baseline AA4 configuration of the SKA Observatory represent a significant step forward in addressing these questions.  With continuum sensitivities at 1.4 GHz and 200 MHz exceeding those of the upgraded GMRT’s band-3 (and band-4), AA4 and AA$^\star$ will be capable of probing average flux densities of order 0.18~$\mu$Jy and 0.21~$\mu$Jy, and 0.60~$\mu$Jy and 0.78~$\mu$Jy in long (= 100 hr) integrations for SKA-Mid and SKA-Low respectively, which is roughly an order of magnitude deeper than current stacking limits ($\sim$1.5 $\mu$Jy). At these levels, it will be possible to distinguish between thermal star-formation-driven emission and weak non-thermal synchrotron signatures from embedded active galaxy, even in the faintest cluster UDGs. Moreover, the combination of wide field-of-view, excellent ($u,v$)-coverage, and arcsecond-scale resolution will enable both statistical studies of large UDG samples and the separation of UDG emission from surrounding diffuse cluster radio halo emission.

\subsection{Expected star-formation rate}

We assume (i) point-source emission, (ii) a spectral index, $\alpha=-0.7$, (iii) the adopted radio--SFR calibration, and (iv) no significant loss of flux density due to resolved-out low-surface-brightness emission.  Clearly, realistic estimates should thus include spectral-index uncertainties, account for effects to the SFR due to cosmic-ray, i.e., systems can under-produce synchrotron emission for a given SFR due to cosmic-ray, and ability to image extended sources faithfully due to limited ($u,v$)-coverage of radio interferometers.

\begin{sidewaysfigure}
\begin{threeparttable}
\centering
\caption{The estimates of 5$\sigma$ detection limits and star-formation rate sensitivities for SKA AA4 and AA$^\star$ configurations.}
\label{tab:ska_sfr_updated}
\begin{tabular}{llcc|ccc|ccc}
\hline
\hline
Telescope & Config.& Frequency & Integration & \textsc{rms} & Flux density  & Equiv.& \multicolumn{3}{c}{SFR$_{5\sigma}$} \\
\cline{8-10}
          &        &       &  time &       & 5$\sigma$ & flux density& Virgo         & Coma           & z = 0.05  \\
          &        &       &       &       &           & 1.4~GHz     & (z = 0.00436) & (z = 0.02316)  &           \\
\cline{5-10}
          &        & (GHz) &  (hr) & \multicolumn{3}{c|}{($\mu$Jy)}  & \multicolumn{3}{c}{($\times$10$^{-4}$ M$_{\odot}$~yr$^{-1}$)}  \\
 \multicolumn{1}{c}{(1)} & \multicolumn{1}{c}{(2)} & \multicolumn{1}{c}{(3)} & \multicolumn{1}{c|}{(4)} & \multicolumn{1}{c}{(5)} & \multicolumn{1}{c}{(6)} & \multicolumn{1}{c|}{(7)} & \multicolumn{1}{c}{(8)} & \multicolumn{1}{c}{(9)} & \multicolumn{1}{c}{(10)} \\
\hline
  SKA-Mid & AA4$^{\rm a}$        & 1.4 & 8$^{\dagger}$ &  0.24 &  1.20 &  1.20 &  0.25 &   9.14 &  44.20 \\
          &                      &     & 100           &  0.07 &  0.35 &  0.35 &  0.07 &   2.67 &  12.90 \\
          & AA$^\star$$^{\rm b}$ &     & 8$^{\dagger}$ &  0.34 &  1.70 &  1.70 &  0.35 &  13.00 &  62.60 \\
          &                      &     & 100           &  0.10 &  0.50 &  0.50 &  0.10 &   3.81 &  18.40 \\
  SKA-Low & AA4$^{\rm c}$        & 0.2 & 8             & 39.44 &197.20 & 50.47 & 10.50 & 385.00 &1860.00 \\
          &                      &     & 100           & 11.85 & 59.25 & 15.17 &  3.16 & 116.00 & 558.00 \\
          & AA$^\star$$^{\rm d}$ &     & 8             & 63.96 &319.80 & 81.87 & 17.00 & 624.00 &3010.00 \\
          &                      &     & 100           & 19.21 & 96.05 & 24.58 &  5.12 & 187.00 & 905.00 \\
\hline
\end{tabular}
\begin{tablenotes}
\footnotesize 
    \item[] We use $A_{\rm eff}/T_{\rm sys}$ for SKA1-Mid from SKAO-TEL-0000818 \citep[Table in Appendix B,][]{Braun2024_SKAO_TEL_0000818} at central frequency = 1.396 GHz ($\approx$ 1.4 GHz), representative continuum bandwidth = 770 MHz (typical SKA-Mid instantaneous usable bandwidth around L-band), no. of polarisations = 2, and natural weighting \citep[see also][as published use case]{Fanaroff2021}.  Below we give system-equivalent flux density (SEFD), and continuum sensitivity (in col.~5) determined using the radiometric equation for the SKA-Mid.
        \item[a] $A_{\rm eff}/T_{\rm sys}$ $= 1751.0$ m$^2$~K$^{-1}$, gives SEFD $= 1.577$ Jy.
        \item[b] $A_{\rm eff}/T_{\rm sys}$ $= 1216.2$ m$^2$~K$^{-1}$, gives SEFD $= 2.270$ Jy.
    \item[] We use central frequency = 200 MHz, continuum bandwidth = 300 MHz (no. of sub-bands = 1), weighting correction factor = 30\% continuum bandwidth and uniform weighting. (We do not use natural weighting here, since it results in poor fits due to large dirty-beam sidelobe levels.) Below we give weighted continuum sensitivity (in col.~5) determined using exposure times calculated with the SKA Sensitivity Calculator at ~\texttt{https://sensitivity-calculator.skao.int} for the SKA-Low.
        \item[c] Using 512 stations, synthesized beam-size = 3.89$^{\prime\prime}$ $\times$ 3.03$^{\prime\prime}$.  The confusion noise, $\sigma_{\rm c}$ = 1.04 $\mu$Jy~beam$^{-1}$.
        \item[d] Using 307 stations, synthesized beam-size = 3.82$^{\prime\prime}$ $\times$ 3.44$^{\prime\prime}$. The confusion noise, $\sigma_{\rm c}$ = 1.21 $\mu$Jy~beam$^{-1}$.
        \item[$\dagger$] We note that in an 8 hr long integration time using 58 dishes (with a maximum baseline of 8 km) of the MeerKAT, the expected thermal noise sensitivity for continuum observations in the L-band (881--1670 MHz) = 3.2~$\mu$Jy~beam$^{-1}$,
        assuming SEFD = 425 Jy, and effective available bandwidth = 385 MHz;  
        from Table 1, available at ~\texttt{https://skaafrica.atlassian.net/wiki/spaces/ESDKB/pages/1486750321/Sensitivity+calculators}. 
        Scaling this to 144 dishes (with a maximum baseline of 36 km) and 197 dishes (with a maximum baseline of 150 km), gives thermal noise sensitivities of 1.28 $\mu$Jy~beam$^{-1}$ and 0.94 $\mu$Jy~beam$^{-1}$, respectively for AA* (interim) and AA4 (design baseline).  Although these estimates are a factor of $\sim$4 higher than our estimates above (col.~5), we believe these are due to improved $A_{\rm eff}/T_{\rm sys}$ and hence improved specifications, $A_{\rm eff}/T_{\rm sys}$ (or SEFD) of the SKA-Mid array.  Additionally, the confusion noise \citep[$\sigma_{\rm c} \propto \theta^2 \alpha^{-0.7}$,][]{Condon1998} would be lower by 2--3 orders of magnitude (or lower by a factor of 40--700).
\end{tablenotes}
\end{threeparttable}
\end{sidewaysfigure}


Table~\ref{tab:ska_sfr_updated} summarises the 5$\sigma$ detection limits and corresponding SFR sensitivities for the SKA-Mid and SKA-Low AA4 and AA$^\star$ configurations for integration times of 8 hr and 100 hr. These predictions indicate that the SKA will be capable of detecting extremely low levels of star formation, on the order of $\sim$10$^{-6}$ M$_{\odot}$ yr$^{-1}$ in UDGs within the Virgo cluster (z = 0.00436), and $\sim$10$^{-3}$--10$^{-1}$~M$_{\odot}$~yr$^{-1}$ for UDGs at z $\approx$ 0.05. With these sensitivities, both AA$^\star$ and especially AA4 should be able to detect the integrated synchrotron emission from UDGs at z $\approx$ 0.05 and beyond. Instead, if such galaxies are truly quenched, the resulting non-detections will yield stringent upper limits on their radio emission, providing key constraints on feedback processes and quenching mechanisms in dense environments. Even if the future looks dark for UDGs, there is still good hope that upcoming facilities, SKA will reveal their true nature.  Hence, AA4 observations are a must to test whether UDGs are predominantly dark-matter–dominated, quiescent systems or whether residual star formation and weak active galaxy still persist in cluster environments.



\section{Discussion}

Deep optical surveys of nearby clusters \citep[e.g., Coma, Virgo:][]{Yagi2016,Mihos2015} find that UDGs occupy the faint, largest, most extended members of the dwarf galaxy population, and that clusters host hundreds of UDGs, many more UDGs than would be predicted from the field population, even after correcting for observational incompleteness and selection biases.  When UDGs in cluster environments are included, they can steepen the measured faint-end of the cluster galaxy luminosity function; note that the exact slope and normalization at absolute magnitude $\gtrsim -12$ remain uncertain because most (spectroscopic) surveys are limited by a lack of understanding as to the selection biases, e.g., surface-brightness completeness, effective (half-light) size cut, area coverage, etc.
Cosmological and zoom simulations predict that the faint end of the UDG abundance is strongly sensitive to host‐halo mass, infall time, and feedback history, and that the lowest-mass systems are strongly affected by tidal effects. In short, the simulations imply that UDGs in clusters are largely quenched, extended dwarfs that are formed via dwarf–environment interactions, while field UDGs may remain gas-rich but diffuse and not centrally concentrated, i.e., low-surface‐density. Such theoretical frameworks help interpret the extremely faint radio and star-formation signatures of UDGs as natural outcomes of their formation scenarios.

\paragraph{UDG abundance} Thus, UDGs have emerged as a key population for probing the link between galaxy formation and environment, particularly within groups and clusters where their number found in a system tends to increase as the mass of the system’s dark-matter halo increases.  Early studies suggested that richer environments contain disproportionately more UDGs, hinting at a possible environmental origin or enhanced survival in massive haloes. 
Recent work by \citet{Makda2022}, using a sample of 51 Hyper Suprime-Cam Subaru identified clusters spanning halo masses from $M_{200} \simeq 0.9\times10^{14}$-- $8.3\times10^{14}$\,M$_\odot$, finds that the abundance of UDGs follows a relatively shallow scaling, $N_{\rm UDG}\propto M_{200}^{0.78\pm0.28}$. This slope is noticeably flatter than the steeper dependence ($N_{\rm UDG}\propto$ M$_{200}^{1.11\pm0.07}$) reported by \citet{vanderBurg2017} based on GAMA groups and clusters. Over the same mass range, the \citet{Makda2022} relation predicts an increase in UDG counts by a factor of about 5 from low- to high-mass clusters, whereas the steeper \citet{vanderBurg2017} relation predicts nearly twice that rise ($\sim$11). 
In other words, the shallower slope may suggest that UDG formation or survival efficiency saturates toward massive haloes, possibly reflecting enhanced disruption, tidal destruction, or completeness limitations at the faint end, and the steeper slope suggest that dense cluster environments promote UDG survival or production more effectively. These contrasting slopes therefore have significant implications for interpreting the UDG population, determining whether they trace halo mass in proportion or instead represent a population whose abundance plateaus in the most massive clusters.

%

Presently, it seems that UDGs represent an extension of the extremely faint, low-luminosity active galactic nucleus population \citep{Ho1997,Ulvestad2001,Ulvestad2002,LalShastri2011} observed in nearby galaxies.
Therefore, similar to low-ionisation nuclear emission-line regions (LINERs) and other weakly active nuclei \cite{Ulvestad2001,Ulvestad2002}, UDGs could host accretion processes operating at very low Eddington ratios ($\lesssim 10^{-5}$--$10^{-3}$ L$_{\rm Edd}$), where radiatively inefficient accretion flows and weak or absent jets produce minimal radio output \citep{LalHo2010,LalShastri2011}. The absence of detectable radio emission from stacked UDG samples is consistent with the faint end of the active galaxy luminosity function \citep{Bontempi2012}, since radio emission may be suppressed by free–free absorption, where the absorber is typically ionised gas (n$_{\rm e}$ = 10$^{-5}$--$10^{-3}$~cm$^{-3}$, T = 10$^4$~K) in the inner narrow-line region or circumnuclear disk, or simply fall below the sensitivity of current instruments. If some UDGs indeed host faint, compact, low-power active galaxy \citep[e.g.,][]{Baldi2019}, their detection would provide valuable constraints on black-hole occupation fractions and feedback efficiency in low-mass, diffuse systems.

%
Note that dust absorption is negligible in the radio band but can play a significant role in the infrared, particularly for UDGs with low-level star formation \citep[see also][]{Magliocchetti2025}. Translating the inferred star-formation rate limits from our radio stacking ($\sim$10$^{-3}$–10$^{-2}$\,M$_\odot$\,yr$^{-1}$) into expected infrared fluxes using the radio–infrared correlation suggests that such UDGs would emit at levels of a few $\mu$Jy at 100–250\,$\mu$m. At the distances of Virgo (16.5\,Mpc) and Coma (100\,Mpc), this corresponds to dust temperatures of roughly 15--25\,K, assuming standard blackbody models, and places them well below current Euclid \citep[][]{Euclid2025} and Herschel \citep[][]{Herschel2010} detection limits. For more distant systems ($z \sim 0.05$), the corresponding fluxes would still be fainter ($<$1\,$\mu$Jy), implying that any dust present in these galaxies must be extremely cold and of very low mass. Thus, even next-generation infrared surveys will struggle to detect the dust emission from such systems unless their star formation is episodically enhanced.

\subsection{Looking ahead}

\begin{table*}
\centering
\caption{Summary of possible causes for extremely weak or undetected radio emission from galaxies at the faint end of the luminosity function, including UDGs and low-luminosity active galaxies.}
\label{tab:summary_lowlum_radio}
\begin{tabular}{p{3.6cm}|p{12cm}}
\hline
\hline
\textbf{Possible cause} & \textbf{Physical explanation and implications} \\
\hline
\textsc{1. Intrinsically weak synchrotron emission} &
Low cosmic-ray electron densities or inefficient particle acceleration in weak active galaxy jets and low star-formation environments lead to faint non-thermal emission. In UDGs, both magnetic field strengths (B $\lesssim 1$--2~$\mu$G) and cosmic-ray densities are likely sub-normal \citep{Hardcastle01.2026.SKA}.  \\[4pt]

\textsc{2. Dominance of thermal (free--free) emission with low normalization} &
In galaxies with extremely low star formation rates ($\lesssim10^{-3}$--$10^{-2}$\,M$_\odot\,\mathrm{yr^{-1}}$), thermal radio emission can dominate but remains below current sensitivity thresholds, yielding no measurable spectral signature \citep{FangxiaAn01.2026.SKA}. \\[4pt]

\textsc{3. Energy losses of cosmic-ray electrons} &
Rapid inverse Compton losses against the cosmic microwave background (especially at higher redshift) and synchrotron cooling in weak magnetic fields can suppress radio flux density at GHz frequencies. The effect is particularly important for diffuse, low-B systems such as UDGs and LINER-like nuclei. \\[4pt]

\textsc{4. Free--free absorption in compact star-forming or active galaxies} &
At low frequencies ($\lesssim$300~MHz), partially ionised gas (e.g., narrow-line region, T $\sim 10^4$ K) can absorb synchrotron emission. Although dust absorption is negligible in radio bands, compact ionised gas or circumnuclear material could obscure low-level radio cores.  Explains many LINER/H\,{\sc ii} and radio-quiet sources \citep{Algera01.2026.SKA,FangxiaAn01.2026.SKA}. \\[4pt]

\textsc{5. Low accretion-rate (radiatively inefficient) active galaxies} &
Many faint LINERs and transition-type nuclei, e.g., H\,{\sc ii} galaxies, radio-quiet sources host radiatively inefficient accretion flows (or advection dominated accretion flows), which produce weak or no observable radio jets. Such mechanisms could explain the absence of radio emission in low-luminosity active galaxies analogs among UDGs. \\[4pt]

\textsc{6. Environmental gas stripping and feedback} &
Cluster UDGs likely experienced strong ram-pressure stripping and tidal interactions that removed cold gas reservoirs, halting both star formation and active galactic nuclei fueling, thereby quenching radio emission. \\[4pt]

\textsc{7. Observational sensitivity limits and surface-brightness dilution} &
For extended, low-surface-brightness galaxies, flux density may be spread over large angular scales, falling below detection thresholds. Deep stacking (median or mean) can recover integrated limits, but single detections remain challenging even with current high-sensitivity instruments \citep{Mazzolari01.2026.SKA}. \\[4pt]

\textsc{8. Intermittent / episodic or stochastic activity cycles} &
Short-lived, recurrent radio episodes in low-luminosity active galaxies may make detection probability small in snapshot surveys. SKA time-domain studies could test this hypothesis. \\[4pt]

\textbf{Towards next-generation (SKA) observations} &
The SKA (AA4 configuration) will deliver $\mu$Jy- to nJy-level sensitivities, enabling direct detections or stringent upper limits for UDGs and faint LINERs out to $z\sim0.05$. Combining SKA continuum and H\,I data will separate gas-rich, star-forming UDGs from quenched, dark systems, constraining the interplay between gas, feedback, and weak radio emission from the galaxy nucleus. \\
\hline
\end{tabular}
\end{table*}

Clearly, UDGs occupy an extreme regime: large effective radii combined with low stellar densities challenge standard models of galaxy formation and feedback. Some UDGs host unexpectedly massive dark matter halos or rich globular cluster systems, indicating diverse formation scenarios. Their low surface brightness and tenuous interstellar medium make them highly sensitive probes of gas stripping, quenching, and star-formation efficiency in dense environments. 
Table \ref{tab:summary_lowlum_radio} summarizes the likely physical causes for the extremely weak or undetected radio emission observed in UDGs and in galaxies at the faint end of the radio luminosity function, such as LINERs, H\,{\sc ii} region galaxies, etc.
The absence of significant synchrotron radiation in these systems may reflect intrinsically weak magnetic fields and inefficient cosmic-ray acceleration, rapid electron energy losses, and free–free absorption or synchrotron self-absorption. Environmental effects such as ram-pressure stripping further quench star formation and active galactic nuclei fueling. Dust plays little role at radio wavelengths, but ionised gas absorption, the low surface brightness extended diffuse emission, and transient low-power active galaxy activity may also limit detection at radio wavelengths. In some cases, radiatively inefficient accretion or free–free absorption within compact ionised regions may further suppress any possible radio emission. Hence, multi-wavelength studies, especially deep radio continuum and H I observations are crucial for revealing residual star formation, cosmic-ray activity, and gas content, providing complementary constraints to optical surveys.

Clearly, UDGs bridge the gap between weak dwarf systems and powerful galaxies, trace the faint end of the galaxy luminosity function, and provide stringent tests of both astrophysical and cosmological models. By studying UDGs, we gain not only a more complete census of galaxies in clusters but also a deeper understanding of the physical processes shaping the low-mass, low-density universe. Their ubiquity, extreme properties, and sensitivity to environment make them indispensable to our understanding of cosmic structure and evolution.

\paragraph{Future studies with the SKA, promise to uncover the faintest regimes of UDG activity}
The cosmological simulations predict clear systematic trends in properties of UDGs. For example, (i) UDGs typically reside in dark-matter halos of $10^{10-11}$~M$_\odot$, corresponding to high mass-to-light ratios (up to $\sim 10^3$ within $R_e$), and show quenched or weakly star-forming stellar populations by $z = 0$. (ii) In high-resolution models, the star-formation rates rarely exceed $10^{-3}$--$10^{-2}$~M$_\odot$~yr$^{-1}$, consistent with a near-quiescent state, and feedback-limited radio flux densities (or infrared fluxes) below current detection thresholds. (iii) The predicted gas fractions decline sharply with increasing environmental density, while rotation-dominated kinematics are more common in field UDGs than in cluster UDGs, which often exhibit inflated velocity dispersions and disturbed morphologies. Moreover, simulated luminosity functions show a flattening of the faint-end slope ($\alpha \approx -1.1$ to $-1.3$) in dense clusters compared to the field ($\alpha \approx -1.5$), reflecting strong environmental quenching.
However, despite recent progress in observations using best, SKA pathfinder and precursor facilities, several key questions, listed below, remain unresolved for UDGs and can be addressed with the SKA, namely SKA continuum and H\,I observations. For example,
\begin{itemize}
\item[$-$] do UDGs host low-luminosity active galactic nuclei or residual star formation? Deep, high-resolution SKA continuum imaging can reveal weak synchrotron cores or diffuse disk emission that are currently below detection limits, testing whether UDGs are truly quiescent or harbor faint, persistent activity.  Thus, distinguish between thermal star-formation-driven emission and weak non-thermal synchrotron signatures from embedded active galaxies, even in the faintest UDGs. \\ [-0.8cm]
\item[$-$] What is the radio–infrared–optical connection in diffuse systems? SKA’s sensitivity across SKA-Low and SKA-Mid will allow combined spectral studies with Euclid and JWST, constraining dust content, magnetic field strength, and the efficiency of cosmic-ray heating in low-surface-brightness extended diffuse galaxies. \\ [-0.8cm]
\item[$-$] How does the cluster environment regulate gas stripping and quenching in UDGs? Observations and models agree that richer halos host proportionally more UDGs, but the normalization and scaling exponent ($N_{\mathrm{UDG}} \propto M_{200}^{x}$) are still being refined \citep[e.g., NUDGEs: $N_{\mathrm{UDG}} \propto M_{200}^{0.78 \pm0.28}$;][]{Makda2022}. Thus, mapping the spatial correlation between UDGs and cluster radio halos or relics can test whether environmental processes such as ram-pressure stripping or tidal interactions suppress the radio emission of UDGs. \\ [-0.8cm]
\item[$-$] Is there a faint-end turnover in the active galactic nuclei luminosity function within clusters? Current luminosity functions may underestimate very low surface-brightness UDGs. Clearly, improved selection tends to increase the number counts of faint sources but uncertainties remain, and different simulations predict different break locations and slopes in the luminosity function.  By probing well below current flux density limits, the SKA will quantify the contribution of UDGs and other low-mass systems to the extremely faint active galaxy population, placing constraints on the fraction of galaxies that host a central supermassive black hole in the low-mass regime. \\ [-0.8cm]
\item[$-$] Can joint SKA continuum and H\,I spectral-line observations disentangle gas-rich, star-forming UDGs from quenched, dark systems? Combining deep continuum maps with future SKA H\,I data will directly link the presence (or absence) of cold gas reservoirs to ongoing star formation and feedback, enabling a clear separation between genuinely dark, gas-poor UDGs and those that retain fuel for intermittent star formation within cluster environments.
\end{itemize}
Together, these investigations will establish whether UDGs represent the low-luminosity, gas-depleted end-state of galaxy evolution or a distinct class of dark-matter-dominated systems with intermittent activity, which is an emerging view from state-of-the-art hydrodynamical simulations \citep{DiCintio2017, Sales2020,Tremmel2020}.
Future deep observations with SKA-Mid and SKA-Low will be crucial to distinguish between genuinely quiescent UDGs and those hosting extremely weak active galactic nuclei activity.

%
%

\section*{Acknowledgments}
We thank the anonymous reviewer for many suggestions that greatly improved the manuscript.
D.V.L. acknowledges the support of the Department of Atomic Energy, Government of India, under project no. 12-R\&D-TFR-5.02-0700.
We thank the staff of the GMRT who have made these observations possible. The GMRT is run by the National Centre for Radio Astrophysics of the Tata Institute of Fundamental Research.

\bibliographystyle{abbrvnat-maxbibnames4}
\bibliography{chapter} 

\begin{thebibliography}{61}
\providecommand{\natexlab}[1]{#1}
\providecommand{\url}[1]{\texttt{#1}}
\expandafter\ifx\csname urlstyle\endcsname\relax
  \providecommand{\doi}[1]{doi: #1}\else
  \providecommand{\doi}{doi: \begingroup \urlstyle{rm}\Url}\fi

\bibitem[Algera et~al.(2026)Algera, author2, author3, author4, and author5]{Algera01.2026.SKA}
H.~S.~B. Algera et al.
\newblock In \emph{Advancing Astrophysics with the SKA -- II (AASKAII)}. 2026.
\newblock arXiv search: Report number AASKAII/Algera01.

\bibitem[An et~al.(2026)An, author2, author3, author4, and author5]{FangxiaAn01.2026.SKA}
F.~X. An et al.
\newblock In \emph{Advancing Astrophysics with the SKA -- II (AASKAII)}. 2026.
\newblock arXiv search: Report number AASKAII/FangxiaAn01.

\bibitem[{Baldi} et~al.(2019){Baldi}, {Capetti}, and {Giovannini}]{Baldi2019}
R.~D. {Baldi}, A.~{Capetti}, and G.~{Giovannini}.
\newblock \emph{\mnras}, 482\penalty0 (2):\penalty0 2294--2304, Jan. 2019.
\newblock \doi{10.1093/mnras/sty2703}.

\bibitem[{Barbosa} et~al.(2020){Barbosa}, {Zaritsky}, {Donnerstein}, {Zhang}, {Dey}, {Mendes de Oliveira}, {Sampedro}, {Molino}, {Costa-Duarte}, {Coelho}, {Cortesi}, {Herpich}, {Hernandez-Jimenez}, {Santos-Silva}, {Pereira}, {Werle}, {Overzier}, {Cid Fernandes}, {Smith Castelli}, {Ribeiro}, {Schoenell}, and {Kanaan}]{Barbosa2020}
C.~E. {Barbosa} et al.
\newblock \emph{\apjs}, 247\penalty0 (2):\penalty0 46, Apr. 2020.
\newblock \doi{10.3847/1538-4365/ab7660}.

\bibitem[{Benavides} et~al.(2021){Benavides}, {Sales}, {Abadi}, {Pillepich}, {Nelson}, {Marinacci}, {Cooper}, {Pakmor}, {Torrey}, {Vogelsberger}, and {Hernquist}]{Benavides2021}
J.~A. {Benavides} et al.
\newblock \emph{Nature Astronomy}, 5:\penalty0 1255--1260, Sept. 2021.
\newblock \doi{10.1038/s41550-021-01458-1}.

\bibitem[{Benavides} et~al.(2023){Benavides}, {Sales}, {Abadi}, {Marinacci}, {Vogelsberger}, and {Hernquist}]{Benavides2023}
J.~A. {Benavides} et al.
\newblock \emph{\mnras}, 522\penalty0 (1):\penalty0 1033--1048, June 2023.
\newblock \doi{10.1093/mnras/stad1053}.

\bibitem[{Bontempi} et~al.(2012){Bontempi}, {Giroletti}, {Panessa}, {Orienti}, and {Doi}]{Bontempi2012}
P.~{Bontempi} et al.
\newblock \emph{\mnras}, 426\penalty0 (1):\penalty0 588--594, Oct. 2012.
\newblock \doi{10.1111/j.1365-2966.2012.21786.x}.

\bibitem[Bourke et~al.(2015)Bourke, Braun, Fender, GOVONI, Green, Hoare, Jarvis, Johnston-Hollitt, Keane, Koopmans, et~al.]{bourke2015advancing}
T.~Bourke et al.
\newblock Advancing astrophysics with the square kilometre array (aaska14), 2015.

\bibitem[{Braun}(2015)]{Braun2015aska.confE.174B}
R.~{Braun}.
\newblock In \emph{Advancing Astrophysics with the Square Kilometre Array (AASKA14)}, page 174, Apr. 2015.
\newblock \doi{10.22323/1.215.0174}.

\bibitem[Braun et~al.(2024)Braun, Bonaldi, Bourke, Cooray, Dewdney, Wagg, et~al.]{Braun2024_SKAO_TEL_0000818}
R.~Braun et al.
\newblock Anticipated ska1 science performance.
\newblock Technical Report SKAO-TEL-0000818, SKA Observatory, 2024.
\newblock Version 2.0, June 2024.

\bibitem[{Carleton} et~al.(2019){Carleton}, {Errani}, {Cooper}, {Kaplinghat}, {Pe{\~n}arrubia}, and {Guo}]{Carleton2019}
T.~{Carleton} et al.
\newblock \emph{\mnras}, 485\penalty0 (1):\penalty0 382--395, May 2019.
\newblock \doi{10.1093/mnras/stz383}.

\bibitem[{Condon} et~al.(1998){Condon}, {Cotton}, {Greisen}, {Yin}, {Perley}, {Taylor}, and {Broderick}]{Condon1998}
J.~J. {Condon} et al.
\newblock \emph{\aj}, 115\penalty0 (5):\penalty0 1693--1716, May 1998.
\newblock \doi{10.1086/300337}.

\bibitem[{Di Cintio} et~al.(2017){Di Cintio}, {Brook}, {Dutton}, {Macci{\`o}}, {Obreja}, and {Dekel}]{DiCintio2017}
A.~{Di Cintio} et al.
\newblock \emph{\mnras}, 466\penalty0 (1):\penalty0 L1--L6, Mar. 2017.
\newblock \doi{10.1093/mnrasl/slw210}.

\bibitem[{Euclid Collaboration} et~al.(2025){Euclid Collaboration}, {Mellier}, {Abdurro'uf}, {Acevedo Barroso}, {Ach{\'u}carro}, {Adamek}, {Adam}, {Addison}, {Aghanim}, {Aguena}, {Ajani}, {Akrami}, {Al-Bahlawan}, {Alavi}, {Albuquerque}, {Alestas}, {Alguero}, {Allaoui}, {Allen}, {Allevato}, {Alonso-Tetilla}, {Altieri}, {Alvarez-Candal}, {Alvi}, {Amara}, {Amendola}, {Amiaux}, {Andika}, {Andreon}, {Andrews}, {Angora}, {Angulo}, {Annibali}, {Anselmi}, {Anselmi}, {Arcari}, {Archidiacono}, {Aric{\`o}}, {Arnaud}, {Arnouts}, {Asgari}, {Asorey}, {Atayde}, {Atek}, {Atrio-Barandela}, {Aubert}, {Aubourg}, {Auphan}, {Auricchio}, {Aussel}, {Aussel}, {Avelino}, {Avgoustidis}, {Avila}, {Awan}, {Azzollini}, {Baccigalupi}, {Bachelet}, {Bacon}, {Baes}, {Bagley}, {Bahr-Kalus}, {Balaguera-Antolinez}, {Balbinot}, {Balcells}, {Baldi}, {Baldry}, {Balestra}, {Ballardini}, {Ballester}, {Balogh}, {Ba{\~n}ados}, {Barbier}, {Bardelli}, {Baron}, {Barreiro}, {Barrena}, {Barriere}, {Barros}, {Barthelemy}, {Bartolo}, {Basset}, {Battaglia},
  {Battisti}, {Baugh}, {Baumont}, {Bazzanini}, {Beaulieu}, {Beckmann}, {Belikov}, {Bel}, {Bellagamba}, {Bella}, {Bellini}, {Benabed}, {Bender}, {Benevento}, {Bennett}, {Benson}, {Bergamini}, {Bermejo-Climent}, {Bernardeau}, {Bertacca}, {Berthe}, {Berthier}, {Bethermin}, {Beutler}, {Bevillon}, {Bhargava}, {Bhatawdekar}, {Bianchi}, {Bisigello}, {Biviano}, {Blake}, {Blanchard}, {Blazek}, {Blot}, {Bosco}, {Bodendorf}, {Boenke}, {B{\"o}hringer}, {Boldrini}, {Bolzonella}, {Bonchi}, {Bonici}, {Bonino}, {Bonino}, {Bonvin}, {Bon}, {Booth}, {Borgani}, {Borlaff}, {Borsato}, {Bose}, {Botticella}, {Boucaud}, {Bouche}, {Boucher}, {Boutigny}, {Bouvard}, {Bouwens}, {Bouy}, {Bowler}, {Bozza}, {Bozzo}, {Branchini}, {Brando}, {Brau-Nogue}, {Brekke}, {Bremer}, {Brescia}, {Breton}, {Brinchmann}, {Brinckmann}, {Brockley-Blatt}, {Brodwin}, {Brouard}, {Brown}, {Bruton}, {Bucko}, {Buddelmeijer}, {Buenadicha}, {Buitrago}, {Burger}, {Burigana}, {Busillo}, {Busonero}, {Cabanac}, {Cabayol-Garcia}, {Cagliari}, {Caillat}, {Caillat},
  {Calabrese}, {Calabro}, {Calderone}, {Calura}, {Camacho Quevedo}, {Camera}, {Campos}, {Ca{\~n}as-Herrera}, {Candini}, {Cantiello}, {Capobianco}, {Cappellaro}, {Cappelluti}, {Cappi}, {Caputi}, {Cara}, {Carbone}, {Cardone}, {Carella}, {Carlberg}, {Carle}, {Carminati}, {Caro}, {Carrasco}, {Carretero}, {Carrilho}, {Carron Duque}, and {Carry}]{Euclid2025}
{Euclid Collaboration} et al.
\newblock \emph{\aap}, 697:\penalty0 A1, May 2025.
\newblock \doi{10.1051/0004-6361/202450810}.

\bibitem[{Fanaroff} et~al.(2021){Fanaroff}, {Lal}, {Venturi}, {Smirnov}, {Bondi}, {Thorat}, {Bester}, {J{\'o}zsa}, {Kleiner}, {Loi}, {Makhathini}, and {White}]{Fanaroff2021}
B.~{Fanaroff} et al.
\newblock \emph{\mnras}, 505\penalty0 (4):\penalty0 6003--6016, Aug. 2021.
\newblock \doi{10.1093/mnras/stab1540}.

\bibitem[{Fielder} et~al.(2024){Fielder}, {Jones}, {Sand}, {Bennet}, {Crnojevi{\'c}}, {Karunakaran}, {Mutlu-Pakdil}, and {Spekkens}]{Fielder2024}
C.~{Fielder} et al.
\newblock \emph{\aj}, 168\penalty0 (5):\penalty0 212, Nov. 2024.
\newblock \doi{10.3847/1538-3881/ad74f6}.

\bibitem[{Gannon} et~al.(2021){Gannon}, {Dullo}, {Forbes}, {Rich}, {Rom{\'a}n}, {Couch}, {Brodie}, {Ferr{\'e}-Mateu}, {Alabi}, and {Mould}]{Gannon2021}
J.~S. {Gannon} et al.
\newblock \emph{\mnras}, 502\penalty0 (3):\penalty0 3144--3157, Apr. 2021.
\newblock \doi{10.1093/mnras/stab277}.

\bibitem[{Gannon} et~al.(2022){Gannon}, {Forbes}, {Romanowsky}, {Ferr{\'e}-Mateu}, {Couch}, {Brodie}, {Huang}, {Janssens}, and {Okabe}]{Gannon2022}
J.~S. {Gannon} et al.
\newblock \emph{\mnras}, 510\penalty0 (1):\penalty0 946--958, Feb. 2022.
\newblock \doi{10.1093/mnras/stab3297}.

\bibitem[{Gannon} et~al.(2024){Gannon}, {Ferr{\'e}-Mateu}, {Forbes}, {Brodie}, {Buzzo}, and {Romanowsky}]{Gannon2024}
J.~S. {Gannon} et al.
\newblock \emph{\mnras}, 531\penalty0 (1):\penalty0 1856--1869, June 2024.
\newblock \doi{10.1093/mnras/stae1287}.

\bibitem[{Gott} et~al.(2001){Gott}, {Vogeley}, {Podariu}, and {Ratra}]{Gott2001}
J.~R. {Gott}, III, M.~S. {Vogeley}, S.~{Podariu}, and B.~{Ratra}.
\newblock \emph{\apj}, 549\penalty0 (1):\penalty0 1--17, Mar. 2001.
\newblock \doi{10.1086/319055}.

\bibitem[Hardcastle et~al.(2026)Hardcastle, author2, author3, author4, and author5]{Hardcastle01.2026.SKA}
M.~J. Hardcastle et al.
\newblock In \emph{Advancing Astrophysics with the SKA -- II (AASKAII)}. 2026.
\newblock arXiv search: Report number AASKAII/Hardcastle01.

\bibitem[{Ho} et~al.(1997){Ho}, {Filippenko}, and {Sargent}]{Ho1997}
L.~C. {Ho}, A.~V. {Filippenko}, and W.~L.~W. {Sargent}.
\newblock \emph{\apj}, 487\penalty0 (2):\penalty0 579--590, Oct. 1997.
\newblock \doi{10.1086/304642}.

\bibitem[{Janssens} et~al.(2019){Janssens}, {Abraham}, {Brodie}, {Forbes}, and {Romanowsky}]{Janssens2019}
S.~R. {Janssens} et al.
\newblock \emph{\apj}, 887\penalty0 (1):\penalty0 92, Dec. 2019.
\newblock \doi{10.3847/1538-4357/ab536c}.

\bibitem[{Jones} et~al.(2021){Jones}, {Bennet}, {Mutlu-Pakdil}, {Sand}, {Spekkens}, {Crnojevi{\'c}}, {Karunakaran}, and {Zaritsky}]{Jones2021}
M.~G. {Jones} et al.
\newblock \emph{\apj}, 919\penalty0 (2):\penalty0 72, Oct. 2021.
\newblock \doi{10.3847/1538-4357/ac0975}.

\bibitem[{Karunakaran} et~al.(2024){Karunakaran}, {Motiwala}, {Spekkens}, {Zaritsky}, {Donnerstein}, and {Dey}]{Karunakaran2024}
A.~{Karunakaran} et al.
\newblock \emph{\apj}, 975\penalty0 (1):\penalty0 91, Nov. 2024.
\newblock \doi{10.3847/1538-4357/ad77cf}.

\bibitem[Koda et~al.(2015)Koda, Yagi, Yamanoi, and Komiyama]{Koda2015}
J.~Koda, M.~Yagi, H.~Yamanoi, and Y.~Komiyama.
\newblock \emph{The Astrophysical Journal Letters}, 807:\penalty0 L2, 2015.
\newblock \doi{10.1088/2041-8205/807/1/L2}.

\bibitem[{Lal} and {Ho}(2010)]{LalHo2010}
D.~V. {Lal} and L.~C. {Ho}.
\newblock \emph{\aj}, 139\penalty0 (3):\penalty0 1089--1105, Mar. 2010.
\newblock \doi{10.1088/0004-6256/139/3/1089}.

\bibitem[{Lal} et~al.(2011){Lal}, {Shastri}, and {Gabuzda}]{LalShastri2011}
D.~V. {Lal}, P.~{Shastri}, and D.~C. {Gabuzda}.
\newblock \emph{\apj}, 731\penalty0 (1):\penalty0 68, Apr. 2011.
\newblock \doi{10.1088/0004-637X/731/1/68}.

\bibitem[{Lal} et~al.(2022){Lal}, {Lyskova}, {Zhang}, {Venturi}, {Forman}, {Jones}, {Churazov}, {van Weeren}, {Bonafede}, {Miller}, {Roberts}, {Bykov}, {Di Mascolo}, {Br{\"u}ggen}, and {Brunetti}]{Lal2022}
D.~V. {Lal} et al.
\newblock \emph{\apj}, 934\penalty0 (2):\penalty0 170, Aug. 2022.
\newblock \doi{10.3847/1538-4357/ac7a9b}.

\bibitem[{Li} et~al.(2023){Li}, {Greene}, {Greco}, {Huang}, {Melchior}, {Beaton}, {Casey}, {Danieli}, {Goulding}, {Joseph}, {Kado-Fong}, {Kim}, and {MacArthur}]{Li2023}
J.~{Li} et al.
\newblock \emph{\apj}, 955\penalty0 (1):\penalty0 1, Sept. 2023.
\newblock \doi{10.3847/1538-4357/ace829}.

\bibitem[{Magliocchetti} et~al.(2025){Magliocchetti}, {La Marca}, {Bisigello}, {Bondi}, {Ricci}, {Fotopoulou}, {Wang}, {Scaramella}, {Pentericci}, {Prandoni}, {Sorce}, {Rottgering}, {Hardcastle}, {Petley}, {La Franca}, {Rubinur}, {Toba}, {Zhong}, {Mezcua}, {Zamorani}, {Shankar}, {Altieri}, {Andreon}, {Auricchio}, {Baccigalupi}, {Baldi}, {Bardelli}, {Biviano}, {Branchini}, {Brescia}, {Brinchmann}, {Camera}, {Canas-Herrera}, {Capobianco}, {Carbone}, {Carretero}, {Castellano}, {Castignani}, {Cavuoti}, {Chambers}, {Cimatti}, {Colodro-Conde}, {Congedo}, {Conselice}, {Conversi}, {Copin}, {Costille}, {Courbin}, {Courtois}, {Cropper}, {Da Silva}, {Degaudenzi}, {De Lucia}, {Di Giorgio}, {Dole}, {Dubath}, {Duncan}, {Dupac}, {Dusini}, {Escoffier}, {Farina}, {Farinelli}, {Faustini}, {Ferriol}, {Finelli}, {Frailis}, {Franceschi}, {Franzetti}, {Fumana}, {Galeotta}, {George}, {Gillis}, {Giocoli}, {Gracia-Carpio}, {Grazian}, {Grupp}, {Haugan}, {Hoar}, {Holmes}, {Hook}, {Hormuth}, {Hornstrup}, {Jahnke}, {Jhabvala},
  {Joachimi}, {Keihanen}, {Kermiche}, {Kiessling}, {Kubik}, {Kummel}, {Kurki-Suonio}, {Le Brun}, {Ligori}, {Lilje}, {Lindholm}, {Lloro}, {Mainetti}, {Maino}, {Maiorano}, {Mansutti}, {Marggraf}, {Martinelli}, {Martinet}, {Marulli}, {Massey}, {Medinaceli}, {Mei}, {Mellier}, {Meneghetti}, {Merlin}, {Meylan}, {Mora}, {Moresco}, {Moscardini}, {Nakajima}, {Neissner}, {Nichol}, {Niemi}, {Padilla}, {Paltani}, {Pasian}, {Pedersen}, {Percival}, {Pettorino}, {Pires}, {Polenta}, {Poncet}, {Popa}, {Pozzetti}, {Raison}, {Renzi}, {Rhodes}, {Riccio}, {Romelli}, {Roncarelli}, {Saglia}, {Sakr}, {Sapone}, {Sartoris}, {Schirmer}, {Schneider}, {Schrabback}, {Secroun}, {Seidel}, {Serrano}, {Simon}, {Sirignano}, {Sirri}, {Stanco}, {Steinwagner}, {Tallada-Crespi}, {Taylor}, {Tereno}, {Tessore}, {Toft}, {Toledo-Moreo}, {Torradeflot}, {Tutusaus}, {Valenziano}, {Valiviita}, {Vassallo}, {Verdoes Kleijn}, {Veropalumbo}, {Wang}, {Weller}, {Zucca}, {Huertas-Company}, and {Scottez}]{Magliocchetti2025}
M.~{Magliocchetti} et al.
\newblock \emph{arXiv e-prints}, art. arXiv:2511.02970, Nov. 2025.
\newblock \doi{10.48550/arXiv.2511.02970}.

\bibitem[{Makda} et~al.(2025){Makda}, {Blyth}, and {Skelton}]{Makda2022}
N.~A. {Makda}, S.-L. {Blyth}, and R.~E. {Skelton}.
\newblock \emph{\mnras}, Oct. 2025.
\newblock \doi{10.1093/mnras/staf1850}.

\bibitem[{Martin} et~al.(2019){Martin}, {Kaviraj}, {Laigle}, {Devriendt}, {Jackson}, {Peirani}, {Dubois}, {Pichon}, and {Slyz}]{Martin2019}
G.~{Martin} et al.
\newblock \emph{\mnras}, 485\penalty0 (1):\penalty0 796--818, May 2019.
\newblock \doi{10.1093/mnras/stz356}.

\bibitem[{Mart{\'\i}nez-Delgado} et~al.(2016){Mart{\'\i}nez-Delgado}, {L{\"a}sker}, {Sharina}, {Toloba}, {Fliri}, {Beaton}, {Valls-Gabaud}, {Karachentsev}, {Chonis}, {Grebel}, {Forbes}, {Romanowsky}, {Gallego-Laborda}, {Teuwen}, {G{\'o}mez-Flechoso}, {Wang}, {Guhathakurta}, {Kaisin}, and {Ho}]{Martinez2016}
D.~{Mart{\'\i}nez-Delgado} et al.
\newblock \emph{\aj}, 151\penalty0 (4):\penalty0 96, Apr. 2016.
\newblock \doi{10.3847/0004-6256/151/4/96}.

\bibitem[Mazzolari et~al.(2026)Mazzolari, author2, author3, author4, and author5]{Mazzolari01.2026.SKA}
G.~Mazzolari et al.
\newblock In \emph{Advancing Astrophysics with the SKA -- II (AASKAII)}. 2026.
\newblock arXiv search: Report number AASKAII/Mazzolari01.

\bibitem[{Mihos} et~al.(2015){Mihos}, {Durrell}, {Ferrarese}, {Feldmeier}, {C{\^o}t{\'e}}, {Peng}, {Harding}, {Liu}, {Gwyn}, and {Cuillandre}]{Mihos2015}
J.~C. {Mihos} et al.
\newblock \emph{\apjl}, 809\penalty0 (2):\penalty0 L21, Aug. 2015.
\newblock \doi{10.1088/2041-8205/809/2/L21}.

\bibitem[{Motiwala} et~al.(2025){Motiwala}, {Karunakaran}, {Spekkens}, {Arora}, {Di Cintio}, {Wright}, {Zaritsky}, and {Macci{\`o}}]{Motiwala2025}
K.~{Motiwala} et al.
\newblock \emph{\apj}, 989\penalty0 (1):\penalty0 86, Aug. 2025.
\newblock \doi{10.3847/1538-4357/ade9a0}.

\bibitem[{Murphy} et~al.(2011){Murphy}, {Condon}, {Schinnerer}, {Kennicutt}, {Calzetti}, {Armus}, {Helou}, {Turner}, {Aniano}, {Beir{\~a}o}, {Bolatto}, {Brandl}, {Croxall}, {Dale}, {Donovan Meyer}, {Draine}, {Engelbracht}, {Hunt}, {Hao}, {Koda}, {Roussel}, {Skibba}, and {Smith}]{Murphy2011}
E.~J. {Murphy} et al.
\newblock \emph{\apj}, 737\penalty0 (2):\penalty0 67, Aug. 2011.
\newblock \doi{10.1088/0004-637X/737/2/67}.

\bibitem[{Papastergis} et~al.(2017){Papastergis}, {Adams}, and {Romanowsky}]{Papastergis2017a}
E.~{Papastergis}, E.~A.~K. {Adams}, and A.~J. {Romanowsky}.
\newblock \emph{\aap}, 601:\penalty0 L10, May 2017.
\newblock \doi{10.1051/0004-6361/201730795}.

\bibitem[{Pilbratt} et~al.(2010){Pilbratt}, {Riedinger}, {Passvogel}, {Crone}, {Doyle}, {Gageur}, {Heras}, {Jewell}, {Metcalfe}, {Ott}, and {Schmidt}]{Herschel2010}
G.~L. {Pilbratt} et al.
\newblock \emph{\aap}, 518:\penalty0 L1, July 2010.
\newblock \doi{10.1051/0004-6361/201014759}.

\bibitem[{Rom{\'a}n} and {Trujillo}(2017{\natexlab{a}})]{Roman2017}
J.~{Rom{\'a}n} and I.~{Trujillo}.
\newblock \emph{\mnras}, 468\penalty0 (1):\penalty0 703--716, June 2017{\natexlab{a}}.
\newblock \doi{10.1093/mnras/stx438}.

\bibitem[{Rom{\'a}n} and {Trujillo}(2017{\natexlab{b}})]{Roman2017a}
J.~{Rom{\'a}n} and I.~{Trujillo}.
\newblock \emph{\mnras}, 468\penalty0 (4):\penalty0 4039--4047, July 2017{\natexlab{b}}.
\newblock \doi{10.1093/mnras/stx694}.

\bibitem[{Rong} et~al.(2017){Rong}, {Guo}, {Gao}, {Liao}, {Xie}, {Puzia}, {Sun}, and {Pan}]{Rong2017}
Y.~{Rong} et al.
\newblock \emph{\mnras}, 470\penalty0 (4):\penalty0 4231--4240, Oct. 2017.
\newblock \doi{10.1093/mnras/stx1440}.

\bibitem[{Sales} et~al.(2020){Sales}, {Navarro}, {Pe{\~n}afiel}, {Peng}, {Lim}, and {Hernquist}]{Sales2020}
L.~V. {Sales} et al.
\newblock \emph{\mnras}, 494\penalty0 (2):\penalty0 1848--1858, May 2020.
\newblock \doi{10.1093/mnras/staa854}.

\bibitem[{Scott} et~al.(2021){Scott}, {Sengupta}, {Lagos}, {Chung}, and {Wong}]{Scott2021}
T.~C. {Scott} et al.
\newblock \emph{\mnras}, 503\penalty0 (3):\penalty0 3953--3964, May 2021.
\newblock \doi{10.1093/mnras/stab390}.

\bibitem[{Shimwell} et~al.(2022){Shimwell}, {Hardcastle}, {Tasse}, {Best}, {R{\"o}ttgering}, {Williams}, {Botteon}, {Drabent}, {Mechev}, {Shulevski}, {van Weeren}, {Bester}, {Br{\"u}ggen}, {Brunetti}, {Callingham}, {Chy{\.z}y}, {Conway}, {Dijkema}, {Duncan}, {de Gasperin}, {Hale}, {Haverkorn}, {Hugo}, {Jackson}, {Mevius}, {Miley}, {Morabito}, {Morganti}, {Offringa}, {Oonk}, {Rafferty}, {Sabater}, {Smith}, {Schwarz}, {Smirnov}, {O'Sullivan}, {Vedantham}, {White}, {Albert}, {Alegre}, {Asabere}, {Bacon}, {Bonafede}, {Bonnassieux}, {Brienza}, {Bilicki}, {Bonato}, {Calistro Rivera}, {Cassano}, {Cochrane}, {Croston}, {Cuciti}, {Dallacasa}, {Danezi}, {Dettmar}, {Di Gennaro}, {Edler}, {En{\ss}lin}, {Emig}, {Franzen}, {Garc{\'\i}a-Vergara}, {Grange}, {G{\"u}rkan}, {Hajduk}, {Heald}, {Heesen}, {Hoang}, {Hoeft}, {Horellou}, {Iacobelli}, {Jamrozy}, {Jeli{\'c}}, {Kondapally}, {Kukreti}, {Kunert-Bajraszewska}, {Magliocchetti}, {Mahatma}, {Ma{\l}ek}, {Mandal}, {Massaro}, {Meyer-Zhao}, {Mingo}, {Mostert}, {Nair},
  {Nakoneczny}, {Nikiel-Wroczy{\'n}ski}, {Orr{\'u}}, {Pajdosz-{\'S}mierciak}, {Pasini}, {Prandoni}, {van Piggelen}, {Rajpurohit}, {Retana-Montenegro}, {Riseley}, {Rowlinson}, {Saxena}, {Schrijvers}, {Sweijen}, {Siewert}, {Timmerman}, {Vaccari}, {Vink}, {West}, {Wo{\l}owska}, {Zhang}, and {Zheng}]{LoTSS-DR2-2022}
T.~W. {Shimwell} et al.
\newblock \emph{\aap}, 659:\penalty0 A1, Mar. 2022.
\newblock \doi{10.1051/0004-6361/202142484}.

\bibitem[{Struble}(2018)]{Struble2018}
M.~F. {Struble}.
\newblock \emph{\mnras}, 473\penalty0 (4):\penalty0 4686--4691, Feb. 2018.
\newblock \doi{10.1093/mnras/stx1785}.

\bibitem[{Tremmel} et~al.(2020){Tremmel}, {Wright}, {Brooks}, {Munshi}, {Nagai}, and {Quinn}]{Tremmel2020}
M.~{Tremmel} et al.
\newblock \emph{\mnras}, 497\penalty0 (3):\penalty0 2786--2810, Sept. 2020.
\newblock \doi{10.1093/mnras/staa2015}.

\bibitem[{Ulvestad} and {Ho}(2001)]{Ulvestad2001}
J.~S. {Ulvestad} and L.~C. {Ho}.
\newblock \emph{\apjl}, 562\penalty0 (2):\penalty0 L133--L136, Dec. 2001.
\newblock \doi{10.1086/338254}.

\bibitem[{Ulvestad} and {Ho}(2002)]{Ulvestad2002}
J.~S. {Ulvestad} and L.~C. {Ho}.
\newblock \emph{\apj}, 581\penalty0 (2):\penalty0 925--931, Dec. 2002.
\newblock \doi{10.1086/344442}.

\bibitem[{van der Burg} et~al.(2016){van der Burg}, {Muzzin}, and {Hoekstra}]{vanderBurg2016}
R.~F.~J. {van der Burg}, A.~{Muzzin}, and H.~{Hoekstra}.
\newblock \emph{\aap}, 590:\penalty0 A20, May 2016.
\newblock \doi{10.1051/0004-6361/201628222}.

\bibitem[{van der Burg} et~al.(2017){van der Burg}, {Hoekstra}, {Muzzin}, {Sif{\'o}n}, {Viola}, {Bremer}, {Brough}, {Driver}, {Erben}, {Heymans}, {Hildebrandt}, {Holwerda}, {Klaes}, {Kuijken}, {McGee}, {Nakajima}, {Napolitano}, {Norberg}, {Taylor}, and {Valentijn}]{vanderBurg2017}
R.~F.~J. {van der Burg} et al.
\newblock \emph{\aap}, 607:\penalty0 A79, Nov. 2017.
\newblock \doi{10.1051/0004-6361/201731335}.

\bibitem[{van Dokkum} et~al.(2022){van Dokkum}, {Shen}, {Keim}, {Trujillo-Gomez}, {Danieli}, {Dutta Chowdhury}, {Abraham}, {Conroy}, {Kruijssen}, {Nagai}, and {Romanowsky}]{vanDokkum2022}
P.~{van Dokkum} et al.
\newblock \emph{\nat}, 605\penalty0 (7910):\penalty0 435--439, May 2022.
\newblock \doi{10.1038/s41586-022-04665-6}.

\bibitem[{van Dokkum} et~al.(2015){van Dokkum}, {Abraham}, {Merritt}, {Zhang}, {Geha}, and {Conroy}]{vanDokkum2015}
P.~G. {van Dokkum} et al.
\newblock \emph{\apjl}, 798\penalty0 (2):\penalty0 L45, Jan. 2015.
\newblock \doi{10.1088/2041-8205/798/2/L45}.

\bibitem[{Van Nest} et~al.(2022){Van Nest}, {Munshi}, {Wright}, {Tremmel}, {Brooks}, {Nagai}, and {Quinn}]{VanNest2022}
J.~D. {Van Nest} et al.
\newblock \emph{\apj}, 926\penalty0 (1):\penalty0 92, Feb. 2022.
\newblock \doi{10.3847/1538-4357/ac43b7}.

\bibitem[{White} et~al.(2007){White}, {Helfand}, {Becker}, {Glikman}, and {de Vries}]{White2007}
R.~L. {White} et al.
\newblock \emph{\apj}, 654\penalty0 (1):\penalty0 99--114, Jan. 2007.
\newblock \doi{10.1086/507700}.

\bibitem[{Yagi} et~al.(2016){Yagi}, {Koda}, {Komiyama}, and {Yamanoi}]{Yagi2016}
M.~{Yagi}, J.~{Koda}, Y.~{Komiyama}, and H.~{Yamanoi}.
\newblock \emph{\apjs}, 225\penalty0 (1):\penalty0 11, July 2016.
\newblock \doi{10.3847/0067-0049/225/1/11}.

\bibitem[{Zaritsky} et~al.(2019){Zaritsky}, {Donnerstein}, {Dey}, {Kadowaki}, {Zhang}, {Karunakaran}, {Mart{\'\i}nez-Delgado}, {Rahman}, and {Spekkens}]{Zaritsky2019}
D.~{Zaritsky} et al.
\newblock \emph{\apjs}, 240\penalty0 (1):\penalty0 1, Jan. 2019.
\newblock \doi{10.3847/1538-4365/aaefe9}.

\bibitem[{Zaritsky} et~al.(2021){Zaritsky}, {Donnerstein}, {Karunakaran}, {Barbosa}, {Dey}, {Kadowaki}, {Spekkens}, and {Zhang}]{Zaritsky2021}
D.~{Zaritsky} et al.
\newblock \emph{\apjs}, 257\penalty0 (2):\penalty0 60, Dec. 2021.
\newblock \doi{10.3847/1538-4365/ac2607}.

\bibitem[{Zheng} et~al.(2025){Zheng}, {Liao}, {Gao}, and {Jiang}]{Zheng2025}
H.~{Zheng}, S.~{Liao}, L.~{Gao}, and F.~{Jiang}.
\newblock \emph{arXiv e-prints}, art. arXiv:2504.14973, Apr. 2025.
\newblock \doi{10.48550/arXiv.2504.14973}.

\bibitem[{Zheng} et~al.(2024){Zheng}, {Faerman}, {Oppenheimer}, {Putman}, {McQuinn}, {Kirby}, {Burchett}, {Telford}, {Werk}, and {Kim}]{Zheng2024}
Y.~{Zheng} et al.
\newblock \emph{\apj}, 960\penalty0 (1):\penalty0 55, Jan. 2024.
\newblock \doi{10.3847/1538-4357/acfe6b}.

\end{thebibliography}

\end{document}